\documentstyle[preprint,aps]{revtex}
\input epsf.tex\def\DESepsf(#1 width #2){\epsfxsize=#2 \epsfbox{#1}}
\begin{document}
\preprint{\vbox{\hbox{}}}
\draft
\title{Some Simple Mixing and Mass Matrices for Neutrinos}
\author{Xiao-Gang He$^a$ and A. Zee$^b$}
\address{
$^a$ Department of Physics, National Taiwan University, Taipei, Taiwan\\
$^b$ Institute for Theoretical Physics, University of California,
Santa Barbara, California 93106, USA}

%
%
\vfill
\maketitle
\begin{abstract}
We argue that the accumulated neutrino data,
including recent results from KamLAND and K2K, point
to a neutrino mixing matrix with
$(V_{11},\; V_{21},\; V_{31}$; $V_{21},\; V_{22},\; V_{32}$;
$V_{13},\; V_{23},\; V_{33})$ = $(-2/\sqrt{6},\; 1/\sqrt{6},\; 1/\sqrt{6}$;
$1/\sqrt{3},\; 1/\sqrt{3},\; 1/\sqrt{3}$; $0,\;
1/\sqrt{2},\; -1/\sqrt{2} )$.
We propose some simple neutrino mass matrices which predict such a
mixing matrix.
\end{abstract}
%
%
%
%
\pagestyle{plain}

\smallskip

In this brief note, we suggest that the accumulating neutrino
data\cite{1,2,3}, including the recent results from KamLAND\cite{4}
and K2K\cite{5}, point to a relatively
simple neutrino mass matrix.
The data can be explained by oscillations between three active 
neutrinos
with the atmospheric neutrino and K2K
data explained by oscillation between the muon
and the tauon neutrinos, and the solar neutrino and KamLAND data explained
by the oscillation between the electron and muon neutrinos.
Following standard convention, let us denote
the neutrinos current eigenstates, coupled to the charged leptons by
the $W$ bosons,
by $\nu _{\alpha }$ ($\alpha =e,$ $\mu ,$ $\tau )$ and the neutrino mass
eigenstates by $\nu _{i}$ ($i=1,2,3).$ We will take the neutrinos to be
Majorana as seems likely\cite{6}, and thus the neutrino mass matrix
$M_{\alpha \beta
}$  is symmetric in the basis of the current eigenstates.
We will also for simplicity assume CP conservation so that $M$ is
real. Thus, $M$ is diagonalized by an orthogonal transformation

\[V^{T}MV=D \]
where the diagonal matrix $D$ has entries $m_{1},m_{2},$ and $m_{3}.$
Clearly, we are free to multiply the Maki-Nakagawa-Sakata\cite{mns} mixing
matrix V on the right by some
diagonal matrix whose diagonal entries are equal to $\pm 1$.
This merely
multiplies each of the columns in V by an arbitrary sign.

We refer the reader to the literature for a detailed analysis of the
data\cite{7,8,9}.
For our purposes, the data may be summarized as follows. Define the mass
squared difference by $\Delta m_{ij}^{2}\equiv $ $m_{i}^{2}-m_{j}^{2}.$
At the 99.3\% confidence level $\Delta
m_{ij}^{2}$ are constrained by
\[1.5\times 10^{-3}eV^{2}\leq |\Delta m_{32}^{2}|
\leq 5.0\times 10^{-3}eV^{2}, \]
and \[ 2.2\times 10^{-5}eV^{2}\leq |\Delta m_{21}^{2}|\leq
2.0\times 10^{-4}eV^{2}, \]
with the best
fit values given by $|\Delta m_{32}^{2}|=3.0\times 10^{-3}$ $eV^{2}$
and $|\Delta m_{21}^{2}|=7.0\times 10^{-5}$ $eV^{2}$. The mixing
angles are in the ranges of $\sin ^{2}2 \theta _{23}>0.85$ and $0.18\leq
\sin^{2}\theta _{12}\leq 0.37$.
Finally, the CHOOZ experiment, done with a French reactor,
failed to see the disappearance of electron antineutrino and thus
gives an upper bound of about $0.22$ on
the $\nu _{e}-\nu _{\tau }$ oscillation parameter
$|V_{13}|$\cite{3}.

We interpret the
data on the mixing angles as follows.
The CHOOZ experiment indicates that
$V_{13}$ is small and so we will simply set it to $0$. We choose,
within the allowed experimental range, $V_{12}\simeq \sin \theta
_{12}\sim 1/\sqrt{3}$ so
that $\tan ^{2}\theta _{12}\sim 1/2.$
This is well within the range $0.37\leq
\tan^{2}\theta _{12}\leq 0.60$ at the 1$\sigma$ level
indicated by the
recent analysis\footnote{ See for example eq. (3.1)
of Bahcall et al. in Ref.\cite{9}.
The bi-maximal matrix gives a too large $\tan\theta_{12} =1$\cite{10}}.
Finally,
the atmospheric neutrino data\cite{2} and K2K data\cite{4}
requires $V_{23}\sim 1/\sqrt{2}.$
In other words, we propose that we know the upper triangle entries of
the matrix $V:$

\[V=\left( \begin{array}{rrr} {X} & {\frac{1}{\sqrt{3}}} & 0 \\
{X} & {X} & {\frac{1}{\sqrt{2}}} \\
{X} & {X} & {X}
\end{array} \right),
\]
where ${X}$ denotes an unknown quantity.

Remarkably, this essentially fixes the
mixing matrix $V$.
Once we take the last column to be
proportional to (0,1,-1), orthogonality and our `` knowledge" that
$|V_{12}|$ is $1/\sqrt{3}$ immediately fix the second column to be
proportional (1,1,1) and hence the first column
to be proportional to (-2,1,1).
We therefore obtain \footnote{Thus V has the pleasing form
that its three columns just correspond
to the three diagonal Gell-Mann matrices of U(3).},

\[V=\left( \begin{array}{rrr}
-{\frac{2}{\sqrt{6}}} & {\frac{1}{\sqrt{3}}} & 0 \\
{\frac{1}{\sqrt{6}}} & {\frac{1}{\sqrt{3}}} & {\frac{1}{\sqrt{2}}} \\
{\frac{1}{\sqrt{6}}} & {\frac{1}{\sqrt{3}}} & {-\frac{1}{\sqrt{2}}}
\end{array} \right). \]
As remarked earlier, we are free to choose the signs of the
column vectors in
the mixing matrix and to make
chiral rotations on the neutrino fields to change the relative signs
of the mass
eigenvalues
\footnote{ Without information on the
relative signs of the eigen-masses, the column vectors can only be
determined up to
$\pm i$. This can
be expressed by multiplying a diagonal
phase matrix $P = Diag(e^{i\sigma}, e^{i\rho}, 1)$ to the right of V.
With CP invariance, $\sigma$ and $\rho$ can take the values of zero
or $\pm \pi/2$.
Neutrinoless double beta decays will provide
some crucial information on these phases.}.

The three column vectors contained in $V$ are the eigenvectors of the matrix

\[
M_{0}=a\left( \begin{array}{rrr}
2 & {0} & 0 \\
{0} & {-1} & {3} \\
{0} & {3} & -1
\end{array} \right),  \]
with eigenvalues $m_{1}=m_{2}=2a,$ and $m_{3}=-4a.$ (The parameter
$a$ merely sets the overall scale.) Thus,
$\Delta m_{21}^{2}=0$ and this pattern  reproduces the data
$|\Delta m_{21}^{2}|/|\Delta m_{32}^{2}|\ll 1$ to first approximation.
Because of the degeneracy in the eigenvalue spectrum, $V$ is not
uniquely determined. We can always replace $V$ by $VW$ where

\[W=\left( \begin{array}{ll}
R & 0 \\
0 & 1
\end{array} \right), \]
with $R$ a $2 \times 2$ rotation matrix. To determine $V,$ and at the
same time to split the degeneracy between $m_{1}\ $and $m_{2},$ we perturb
$M_{0}$ to $M=M_{0}+\delta M_T,$ where

\[\delta M_T=\varepsilon a\left( \begin{array}{lll}
0 & 1 & 1 \\
1 & 0 & 1 \\
1 & 1 & 0
\end{array}
\right).
\]

We have the mass eigenvalues
$m_{1}=2a(1-\varepsilon/2 ),m_{2}=2a(1+\varepsilon ),$ and
$m_{3}=-4a(1+\varepsilon/4).$ Thus, to the lowest order, we can determine
$\varepsilon = \Delta m_{21}^{2}/\Delta m_{32}^{2}$.
The overall scale of the mass matrix $a$ is given by
$a^2 = \Delta m^2_{32}/12$.

Note that our proposed neutrino Majorana mass matrix $M$ is traceless.
One may be tempted to conjecture that this property may provide a clue to
the origin of the mass matrix $M$.
As is well known, a general Majorana matrix for the neutrinos has 9
real parameters while feasible experiments can measure only 7 of these. It has
been suggested that conditions such as
$Det M =0$\cite{11},
texture zeros or other
relations\cite{12,13,14,15,16}
be imposed to cut down on
the number of parameters. Our example here satisfies $Tr M = 0$, but not
$Det M=0$. In a forthcoming paper\cite{17},
we give a phenomenological analysis of
the data imposing the condition $Tr M = 0$, which is generally
satisfied by models in which $M$ is given by the
commutator\footnote{An example is the simplest version of
the many so-called Zee models\cite{11}.} of two
matrices $M=[A,B]$.

Other perturbations
can also lead to the
same mixing matrix $V$ while splitting the degeneracy $\Delta
m_{21}^{2}=0$. An interesting example is the
`` democratic'' form

\[\delta M_D=\varepsilon a\left( \begin{array}{lll}
1 & 1 & 1 \\
1 & 1 & 1 \\
1 & 1 & 1
\end{array}
\right).
\]

The matrix $\delta M_D$ is evidently a projection matrix that projects the
first and third columns in V to zero.
Thus, the eigenvalues are given by
$m_{1}=2a,m_{2}=2a(1+3\varepsilon/2 ),$ and
$m_{3}=-4a.$
Again $\varepsilon $ and $a^2$ are given by, to the lowest order,
$\varepsilon = \Delta m_{21}^{2}/\Delta m_{32}^{2}$ and
$a^2 = \Delta m^2_{32} /12$, respectively.
We note that this mass matrix is not traceless.

We mention that there is a whole class of models we can propose.
Generalize $M_0$ to be

\[ \tilde M_0= a\left( \begin{array}{ccc}
2 & 0 & 0 \\
0 & 1-y & 1+y \\
0 & 1+y & 1-y
\end{array}
\right),
\]
with the case mentioned earlier corresponding to y=2.
Thus in general we propose

\[ M=\tilde M_0+  \delta M, \]
with $\delta M$ being $\delta M_T$ or $\delta M_D$. They lead to the
same mixing matrix $V$, with the eigenvalues $m_i$ given by
$(2a[1-\varepsilon /2], 2a[1+\varepsilon], -2a[y +\varepsilon/2])$ and
$(2a, 2a[1+3\varepsilon /2], -2ay)$, respectively.

Note that the most general mass matrix which produces the
mixing matrix V
can be expressed as linear combinations of the three matrices of the forms
given by $M_0$, $\delta M_T$ and $\delta M_D$. Once we
committed to a specific form for $M$, the three parameters specifying the
linear combination merely parameterize the three neutrino eigen-masses
$m_{1,2,3}$.  Also for any given mixing
matrix, the mass matrix can be
specified by mass eigenvalues.

Our purpose here is evidently not to give a detailed fit to the data,
but to suggest some relatively simple and appealing mass matrices. The
appearance of simple integers in the mixing and mass matrices we proposed
is perhaps intriguing and provides a glimmer of a hope that they may
be obtained by group theoretic considerations. To provide a
theoretical origin of the mass matrix $M$ presents an interesting
challenge.

\newpage
\noindent
{\bf\large Acknowledgments}

This work
was supported in part by
NSC under grant number NSC
91-2112-M-002-42,
and by the MOE Academic Excellence Project 89-N-FA01-1-4-3 of Taiwan,
and by NSF under grant number PHY 99-07949 of USA.
AZ thanks Professor Pauchy Hwang and the Department of Physics of the
National Taiwan University, where this work was initiated,
for warm hospitality.
\\
\\
{\bf Note Added}

It has been called to our attention that the mixing matrix $V$ we
suggested was proposed by L. Wolfenstein\cite{wol} more than 20 years 
ago (but with the first and second column interchanged), and by
P.F. Harrison, D.H. Perkins and W.G. Scott\cite{18} before the SNO and
KamLAND data came out, and had been studied by them and by Z.-Z. Xing
\cite{19,20}. Our discussion
of the mass matrix, however, seems to be novel. The mixing
matrix $V$ is a special case of a family of mixing matrices obtained by
C.S. Lam \cite{21} by imposing ``2-3 symmetry''. He fixed $V$ further by
fitting it to the data available at the time.


\begin{thebibliography}{99}

\bibitem{1}

Q.R. Ahmad et al., (SNO Collaboration), Phys. Rev. Lett. {\bf 89},
011301(2002);
Phys. Rev. Lett. {\bf 89}, 011302(2002);
S. Fukuda et al., (Super-Kamiokande Collaboration),
Phys. Lett. {\bf B539}, 179(2002); B.T. Cleveland et al., Astrophys.
{\bf J. 496}, 505(1998); R. Davis, Prog. Part. Nucl. Nucl. Phys.
{\bf 32}, 13(1994); D. N. Abdurashitov et al., (SAGE Collaboration),
Phys. Rev. {\bf D60}, 055801(1999); arXiv: astro-ph/0204245;
W. Hampel et al., (GALLEX Collaboration), Phys. Let. {\bf B447}, 127(1999);
C. Cattadori, (GNO Collaboration), Nucl. Phys. {\bf B111} (Proc. Suppl.),
311(2002).

\bibitem{2} Y. Fukuda et al., (Kamiokande Collaboration), Phys. Lett.
{\bf B335}, 237(1994); R. Becker-Szendy et al., (IMB Collaboration),
Nucl. Phys. {\bf B38} (Proc. Suppl.), 331(1995); W.W.M. Allison et al.,
(Soudan Collaboration) Phys. Lett. {\bf B449}, 137(1999);
M. Ambrosio et al., (MACRO Collaboration) Phys. Lett. {\bf B434}, 451(1998);
M. Soili, arXiv: hep-ex/0211030.


\bibitem{3} M. Apollonio et al., (CHOOZ Collaboration), Phys. Lett.
{\bf 466}, 415(1999) [arXiv: hep-ph/9907037].

\bibitem{4} K. Eguchi et al., (KamLAND Collaboration), arXiv:hep-ph/0212021.

\bibitem{5} M.H. Ahn et al., (K2K Collaboration), arXiv:hep-ph/0212007.

\bibitem{6} For a review, see for example, B. Kayser, arXiv: hep-ph/0211134.

\bibitem{mns} Z. Maki, M. Nakagawa and S. Sakata, Prog. Theor. Phys. 
{\bf 28}, 870(1962).

\bibitem{7} V. Barger et al., Phys. Lett. {\bf B537},179(2002) [arXiv:
hep-ph/0204253]; J.N. Bachall, M.C. Gonzalez-Garcia and C. Pena-Garay,
JHEP {\bf 0207}, 054(2002) [arXiv: hep-ph0204314];
M. Maltoni et al., arXiv: hep-ph/0207227;
G.L. Fogli et al., arXiv: hep-ph/0208026.

\bibitem{8} 
P. Creminelli, G. Signorelli and A. Strumia, {\bf JHEP} 0105:052(2001)
(Addendum)
[arXiv: hep-ph/0102234];
V. Barger and D. Marfatia, arXiv: hep-ph/0212126;
G. Fogli et al., arXiv: hep-ph/0212127;
M. Maltoni, T. Schwetz and J. Valle, arXiv: hep-ph/0212129;
A. Bandyopadhyay et al., arXiv: hep-ph/0212146;
H. Nunokawa and W. Teves and R. Z. Funchal,
hep-ph/0212202; P. Aliani et al., arXiv: hep-ph0212212;
P.C. de Holanda and A. Yu. Smirnov,
arXiv hep-ph/0212270; S. Pakvasa and J. Valle, arXiv: hep-ph/0301061;
A. Balantekin and H. Yuksel, arXiv: hep-ph/0301072.

\bibitem{9} J.N. Bachall, M.C. Gonzalez-Garcia and C. Pena-Garay,
arXiv: hep-ph/0212147.

\bibitem{10} V. Barger, S. Pakvasa, T. Weiler and K. Whisnant, 
Phys. Lett. {\bf B437}, 107(1998).

\bibitem{11} G.C. Branco, R. Felipe, F. Joaquim and T. Yanagida, 
arXiv: hep-ph/0212341.

\bibitem{12} A. Zee, Phys. Lett. {\bf 93B}, 389(1980); {\bf 161B},
141(1985); L. Wolfenstein, Nucl. Phys. {\bf B175}, 93(1980); 
For a brief review, see  D. Chang and A. Zee, Phys. Rev.
{\bf D61}, 071303(2000).

\bibitem{13} Y. Smirnov and Zhijian Tao, Nucl. Phys. {\bf B426}, 415(1994);
Y. Smirnov and M. Tanimoto, Phys. Rev. {\bf D55}, 1665(1997);
C. Jarlskog et al., Phys. Lett. {\bf B449}, 240(1999);
P. Frampton and S. Glashow, Phys. Lett. {\bf B461}, 95(1999).

\bibitem{14}
Y. Koide, Phys. Rev. {\bf D64}, 077301(2001);
P. Frampton, M. Oh and T. Yoshikawa, Phys. Rev. {\bf D65}, 073014(2002).

\bibitem{15}
A. Joshipura and S. Rindani, Phys. Lett. {\bf B464}, 239(1999);
K.-M. Cheung and O. Kong, Phys. Rev. {\bf D16}, 113012(2000);
K. Baliji, W. Grimus and T. Schwetz, Phys. Lett. {\bf B508}, 301(2001);
E. Mitsuda and K. Sasaki, Phys. Lett. {\bf B516}, 47(2001);
A. Ghosal, Y. Koide and H. Fusaoka, Phys. Rev. {\bf D64}, 053012(2001);
D. Dicus, H.-J. He and J. Ng, Phys. Rev. Lett. {\bf 87}, 111803(2001);
B. Brahmachari and S. Choubey, Phys. Lett. {\bf B531}, 99(2002);
T. Kitabayashi and M. Yasue, Int.J. Mod. Phys. {\bf A17}, 2519(2002);
M.-Y. Cheng and K. Cheung, arXiv: hep-ph/0203051;
Y. Koide, Nucl. Phys. Proc. Suppl. {\bf 111}, 294(2002).

\bibitem{16} P. H. Frampton, S.L. Glashow and D. Marfatia,
Phys. Lett. {\bf B 536}, 79(2002) [arXiv: hep-ph/0202008];
M.-C. Chen and K.T. Mahanthappa, arXiv: hep-ph/0212375;
B. Desai, D. Roy and A, Vaucher, arXiv: hep-ph/02035;
W.-L. Guo and Z.-Z. Xing, arXiv: hep-ph/0212142;
Z.-Z. Xing, arXiv: hep-ph/0210276;
E. Ma, Phys. Rev. {\bf D66}, 117301(2002);
A. Kageyama et al., Phys. Rev. {\bf D65}, 096010(2002);
S. Barr, J. Int. Mod. Phys. {\bf A16S1B}, 579(2001);
I. Dorsner and S. Barr, Nucl. Phys. {\bf B617}, 493(2001);
A.J. Davies and X.-G. He, Phys. Rev. {\bf D46},
3208(1992); K. Babu and Q. Shfi, Phys. Lett. {\bf B311}, 172(1993);
J. H. Hewett and T.G. Rizzo, Phys. Rev.
{\bf D33}, 1519(1996).

\bibitem{17} X.-G. He and A. Zee, in preparation.

\bibitem{wol} L. Wolfenstein, Phys. Rev. {\bf D18}, 958(1978).

\bibitem{18} P.F. Harrison, D. H. Perkins and W.G. Scott,
Phys. Lett. {\bf B458}, 79(1999).

\bibitem{19}
W.G. Scott, Nucl. Phys. Proc. Suppl. {\bf 85}, 177(2000)
[arXiv: hep-ph/9909431];
P. F. Harrison, D.H. Perkins and W.G. Scott,
Phys. Lett. {\bf B530}, 167(2002) [arXiv: hep-ph/0202074];
P. F. Harrison and W.G. Scott, Phys. Lett. {\bf B535}, 163(2002);
arXiv: hep-ph/0302025.

\bibitem{20}
Z.-Z. Xing, Phys. Lett. {\bf B533}, 85(2002) [arXiv:
hep-ph/0204049].

\bibitem{21} C.S. Lam Phys. Lett. {\bf B507}, 214(2002)
[arXiv: hep-ph/0104116].

\end{thebibliography}
\end{document}